\documentstyle[12pt,emlines]{article}
\newcommand{\dfrac}{\displaystyle \frac}
\newcommand{\Dsla}{D\hspace{-9.25pt}  /}
\newcommand{\nb}[1]{\mbox{\normalsize #1}}
\def\x{\hspace{1ex}}
\def\ao{{}\kern-.10em\hbox{``}}
\begin{document}
\large
\bibliographystyle{plain}

\begin{titlepage}
\hfill \begin{tabular}{l}
            HEPHY-PUB 593/93\\
            UWThPh-1993-56\\
            November 1993
       \end{tabular}\\[4cm]
\begin{center}
{\Large \bf COMPLETE CLASSES OF GUTS WITH VANISHING ONE-LOOP BETA
FUNCTIONS}\\
\vspace{1.5cm}
{\Large \bf Wolfgang LUCHA}\\[.5cm]
{\large Institut f\"ur Hochenergiephysik\\
\"Osterreichische Akademie der Wissenschaften\\
Nikolsdorfergasse 18, A-1050 Wien, Austria}\\[1cm]
{\Large \bf Franz F. SCH\"OBERL}\\[.5cm]
{\large Institut f\"ur Theoretische Physik\\
Universit\"at Wien\\
Boltzmanngasse 5, A-1090 Wien, Austria}\\[3cm]
{\bf Abstract}
\end{center}
\normalsize

\noindent
By explicit solution of the one-loop finiteness conditions for all
dimensionless coupling constants (i.~e., gauge coupling constant as well as
Yukawa and quartic scalar-boson self-interaction coupling constants), two
classes of grand unified theories characterized by renormalization-group beta
functions which all vanish at least at the one-loop level are constructed and
analyzed with respect to the (suspected) appearance of quadratic divergences,
with the result that without exception in all of these models the masses of
both vector and scalar bosons receive quadratically divergent one-loop
contributions.
\end{titlepage}

\section{Introduction}

Supersymmetry---apart from the important fact of being the only one among the
possible symmetries of a nontrivial S-operator \cite{haag75} which up to now
has not been discovered experimentally---attracts continuous interest because
of its far-reaching ability of softening the high-energy behaviour of quantum
field theories. $N = 1$ supersymmetric theories satisfying two so-called \ao
finiteness conditions" are finite at least up to two loops \cite{N=1}, even
if softly broken \cite{N=1softbreak}. $N = 2$ supersymmetric theories
satisfying just a single one-loop finiteness condition prove to be finite at
all orders of their perturbation expansion \cite{howe83a}, and this once
again even in case of being softly broken \cite{N=2softbreak}.
The most famous representative of this latter class of finite quantum field
theories \cite{N=2class} is the well-known $N = 4$ super-Yang--Mills theory
where the single finiteness condition is automatically satisfied
\cite{N=4SYM}.

It is no wonder that soon after the discovery of the supersymmetric finite
quantum field theories some in a certain sense \ao inverse" questions have
been put forward \cite{lucha86a,lucha86b,lucha87a}: Of what kind are the
consequences of the requirement of at least perturbative finiteness for the
most general renormalizable quantum field theory? Is, in particular,
supersymmetry indeed a necessary prerequisite for finiteness? Do there exist
any other, non-supersymmetric finite theories \cite{lucha87b}?

In the sequel, some attempts have been undertaken to enlarge the class of
finite quantum field theories by building non-supersymmetric finite models.
For instance, in Ref. \cite{shapiro93} two sets of non-supersymmetric grand
unified theories have been proposed which are characterized by the demand of
vanishing one-loop renormalization-group $\beta$ functions for all, in any
case dimensionless coupling constants of the theory, and explicit models
corresponding to the two or three of the allowed gauge groups with smallest
dimension have been given.\footnote{\normalsize\ In view of our findings the
models of Ref. \cite{shapiro93} do not represent the whole truth.}

The main aim of the actual investigation is to analyze the finiteness
conditions resulting from the above requirement of vanishing one-loop beta
functions by (as far as manageable) algebraic methods in order to determine
the complete classes of the corresponding theories. After a brief recall, in
Sect. \ref{sec:twomods}, of the definition of these two classes of models we
discuss their finiteness conditions, first, in Sect. \ref{sec:simplsol}, for
the presumably simpler case and then, in Sect. \ref{sec:gensol}, for the
certainly more delicate case, and summarize our conclusions in Sect.
\ref{sec:solconc}.

\section{Two Models with Vanishing One-Loop Beta
Functions}\label{sec:twomods}

The two models \cite{shapiro93}---or, more precisely, two classes of
models---with vanishing one-loop contributions to the renormalization-group
beta functions of all of their dimensionless coupling constants, which are to
be considered in the present investigation, are characterized by three main
features:
\begin{enumerate}
\item Their gauge group ${\cal G}$ is assumed to be some special unitary
group SU($N$): ${\cal G} = \mbox{SU($N$)}$.
\item Their particle content is assumed to involve only particles which
transform either according to the fundamental representation or according to
the adjoint representation of ${\cal G}$.
\item Their Lagrangian does not involve any dimensional parameter.
\end{enumerate}
The primary advantage of this very specific and simple choice is that in both
of these models all (then necessarily dimensionless) couplings may be
expressed solely in terms of
\begin{itemize}
\item the generators $T^a$, with $a = 1,2,\dots , N^2 - 1$, in the {\em
fundamental\/} representation of ${\cal G}$, the normalization of which is
fixed by their second-order Dynkin index $T_{\nb{f}}$, defined by $T_{\nb{f}}
\, \delta_{ab} := \mbox{Tr}(T^a \, T^b)$;
\item the generators $\frac{1}{i} \, f_{abc}$ in the {\em adjoint\/}
representation of ${\cal G}$, where $f_{abc}$ denote the (completely
antisymmetric) structure constants which define the gauge group under
consideration; or
\item the completely symmetric constants
$$
d_{abc} \equiv
\dfrac{\mbox{Tr}\left(\left\{T^a,T^b\right\}T^c\right)}{T_{\nb{f}}} \quad .
$$
\end{itemize}
The postulated gauge invariance requires, of course, both models to contain
(real) gauge vector bosons $V_\mu^a$ in the adjoint representation of ${\cal
G}$, which enter in the field strength $F_{\mu\nu}^a \equiv \partial_\mu
V_\nu^a - \partial_\nu V_\mu^a + g \, f_{abc} \, V_\mu^b \, V_\nu^c$ as well
as in the covariant derivative $D_\mu \equiv \partial_\mu - i \, g \, V_\mu^a
\, T^a$.

\subsection{The general model}\label{subsec:genmod}

Apart from the above-mentioned gauge bosons $V_\mu^a$, the particle content
of this model consists of
\begin{itemize}
\item $m$ sets of Dirac fermions $\Psi_{(k)}$, $k = 1,2,\dots ,m$, each of
these sets in the adjoint representation of ${\cal G}$;
\item $m$ sets of Dirac fermions $\chi_{(k)}$, $k = 1,2,\dots ,m$, each of
these sets in the fundamental representation of ${\cal G}$;
\item $n$ sets of Dirac fermions $\zeta_{(k)}$, $k = 1,2,\dots ,n$, each of
these sets in the fundamental representation of ${\cal G}$;
\item real scalar bosons $\Phi$ in the adjoint representation of ${\cal G}$;
\item complex scalar bosons $\varphi$ in the fundamental representation of
${\cal G}$.
\end{itemize}
This general model is defined by the Lagrangian \cite{shapiro93}
\begin{eqnarray}
{\cal L}
&=& - \dfrac{1}{4} \, F_{\mu\nu}^a \, F^{\mu\nu}_a
+ i \sum_{k = 1}^m \bar\Psi_{(k)}^a \left( \Dsla_{ab} - h_1 \, f_{abc} \,
\Phi^c \right) \Psi_{(k)}^b \nonumber\\
&+& i \sum_{k = 1}^m \bar\chi_{(k)} \left( \Dsla - i \, h_2 \, T^a \, \Phi^a
\right) \chi_{(k)}
+ i \sum_{k = 1}^n \bar\zeta_{(k)} \, \Dsla \: \, \zeta_{(k)} \nonumber\\
&+& \left( i \, h_3 \sum_{k = 1}^m \bar\chi_{(k)} \, T^a \, \varphi \,
\Psi_{(k)}^a + \mbox{H. c.} \right) \nonumber\\
&+& \dfrac{1}{2} \left( D_\mu \Phi \right)^T D^\mu \Phi
+ \left( D_\mu \varphi \right)^\dagger D^\mu \varphi
- \dfrac{\lambda_1}{8} \left( \Phi^T \Phi \right)^2
- \dfrac{\lambda_2}{8} \left( \Phi^a \, d_{abc} \, \Phi^b \right)^2
\nonumber\\
&-& \dfrac{\lambda_3}{2} \left( \Phi^T \Phi \right) \left( \varphi^\dagger
\varphi \right)
- \dfrac{\lambda_4}{2} \left( \Phi^a \, d_{abc} \, \Phi^b \right) \left(
\varphi^\dagger \, T^c \, \varphi \right)
- \dfrac{\lambda_5}{2} \left( \varphi^\dagger \varphi \right)^2 \quad .
\label{eq:Lagr-genmod}
\end{eqnarray}
The fermions $\chi_{(k)}$ are discriminated from the fermions $\zeta_{(k)}$
by the fact that the former also undergo Yukawa interactions whereas the
latter do not. In addition to the gauge coupling constant $g$, this general
model involves three Yukawa coupling constants, $h_1,h_2,h_3$, and five
scalar-boson self-coupling constants, $\lambda_1,\lambda_2,\dots ,\lambda_5$.
Our aim will be to investigate the consequences of the required vanishing of
the one-loop beta functions for these coupling constants.

The finiteness condition which may be satisfied most easily is the one for
the one-loop contribution to the renormalization of the gauge coupling
constant \cite{shapiro93,lucha93a}:
\begin{equation}
21 \, N - 4 \, [(2 \, N + 1) \, m + n] = 1 \quad .
\label{eq:olf-gc/gen}
\end{equation}
It obviously restricts
\begin{itemize}
\item the possible gauge groups SU($N$) to the values $N = 4 \, \ell + 1$
for $\ell = 1,2,\dots$, i.~e., to the values $N = 5,9,13,\dots$, and
\item the multiplicities $m$ and $n$ to three possible combinations: from the
positivity of the multiplicity $n$, i.~e., $n \ge 0$, the multiplicity $m$ is
bounded from above by
$$
m \le \dfrac{21 \, N - 1}{4 \, (2 \, N + 1)} < 3
\quad \mbox{for arbitrary $N > 0$} \quad ,
$$
which restricts $m$ to the values $m = 0,1,2$, the corresponding values of
$n$ then being fixed by Eq. (\ref{eq:olf-gc/gen}).
\end{itemize}

\subsection{The simplified model}

This model is obtained from the more general model described above by
completely decoupling the fermions $\Psi_{(k)}$ and $\zeta_{(k)}$ as well as
the scalar bosons $\varphi$ from the theory. Accordingly, the non-vector
particle content of this model consists of
\begin{itemize}
\item $m$ sets of Dirac fermions $\chi_{(k)}$, $k = 1,2,\dots ,m$, each of
these sets in the fundamental representation of ${\cal G}$;
\item real scalar bosons $\Phi$ in the adjoint representation of ${\cal G}$.
\end{itemize}
Consequently, this simplified model is defined by the Lagrangian
\cite{lucha93a}
\begin{eqnarray}
{\cal L}
&=& - \dfrac{1}{4} \, F_{\mu\nu}^a \, F^{\mu\nu}_a
+ i \sum_{k = 1}^m \bar\chi_{(k)} \left( \Dsla - i \, h \, T^a \, \Phi^a
\right) \chi_{(k)} \nonumber\\
&+& \dfrac{1}{2} \left( D_\mu \Phi \right)^T D^\mu \Phi
- \dfrac{\lambda_1}{8} \left( \Phi^T \Phi \right)^2
- \dfrac{\lambda_2}{8} \left( \Phi^a \, d_{abc} \, \Phi^b \right)^2 \quad .
\label{eq:simplmod-lagr}
\end{eqnarray}
It now involves only one Yukawa coupling constant, $h$ (the former $h_2$),
and only two scalar-boson self-coupling constants, $\lambda_1,\lambda_2$.

Finiteness of the one-loop contribution to the renormalization of the gauge
coupling constant, as expressed here by the relation
\cite{shapiro93,lucha93a}
\begin{equation}
21 \, N = 4 \, m \quad ,
\label{eq:olf-gc/simpl}
\end{equation}
now restricts the possible gauge groups SU($N$) to the values $N = 4 \, \ell$
for $\ell = 1,2,\dots$, i.~e., to the values $N = 4,8,12,\dots$.

According to the finiteness condition (\ref{eq:olf-gc/simpl}), for any
simplified model with vanishing one-loop gauge-coupling beta function there
exists a unique relation between the considered gauge group, i.~e., the value
of $N$, and the allowed multiplicity $m$. In the subsequent discussion this
relation will be taken into account already from the very beginning by
introducing, in place of $N$ and $m$, the \ao group parameter"
$$
\ell = \dfrac{N}{4} = \dfrac{m}{21} \quad .
$$

\section{Solutions for the Simplified Model}\label{sec:simplsol}

For the simplified model, the conditions for the one-loop contributions to
the renormalization-group beta functions of the Yukawa coupling $h$ and the
quartic scalar-boson self-couplings $\lambda_i$, $i = 1,2$, to vanish
\cite{shapiro93}, expressed in terms of the three real and non-negative
variables
$$
x \equiv \dfrac{h^2}{g^2} \ge 0
$$
and
$$
y_i \equiv \dfrac{\lambda_i}{g^2} \ge 0 \quad , \qquad i = 1,2 \quad ,
$$
read for arbitrary gauge groups SU($N$) with $N = 4 \, \ell$, $\ell =
1,2,\dots$:
\begin{eqnarray}
&&\left( 184 \, \ell^2 - 3 \right) x^2
- 6 \left( 16 \, \ell^2 - 1 \right) x = 0 \quad ,
\label{eq:simplfc(1)}\\[1ex]
&&\left( 16 \, \ell^2 + 7 \right) y_1^2 + 24 - 48 \, \ell \, y_1
+ 84 \, \ell \, x \, y_1 \nonumber\\
&&\qquad + \: \dfrac{16 \, \ell^2 - 4}{\ell}
\left( y_1 \, y_2 + \dfrac{1}{2 \, \ell} \, y_2^2 \right)
- 21 \, x^2 = 0 \quad ,
\label{eq:simplfc(2)}\\[1ex]
&&\dfrac{16 \, \ell^2 - 15}{\ell} \, y_2^2 + 12 \, \ell - 48 \, \ell \, y_2
+ 12 \, y_1 \, y_2 + 84 \, \ell \, x \, y_2 - 42 \, \ell \, x^2 \nonumber\\
&&\qquad = 0 \quad .
\label{eq:simplfc(3)}
\end{eqnarray}
Quite obviously, Eq. (\ref{eq:simplfc(1)}) allows for two and only two
solutions for the variable $x$, viz.
\begin{equation}
x = 0
\label{eq:simpl-xsol1}
\end{equation}
or
\begin{equation}
x = \dfrac{6 \left( 16 \, \ell^2 - 1 \right)}{184 \, \ell^2 - 3} \quad .
\label{eq:simpl-xsol2}
\end{equation}

Our first step in the course of determination of the complete set of
solutions to the set of equations (\ref{eq:simplfc(1)}) to
(\ref{eq:simplfc(3)}) given above is the proof of the following statement:
{\em For the second, i.~e., non-vanishing, solution (\ref{eq:simpl-xsol2}) of
Eq. (\ref{eq:simplfc(1)}) for the variable $x$ and for arbitrary values of
the group parameter $\ell = 1,2,\dots$, the remaining set of equations
(\ref{eq:simplfc(2)}), (\ref{eq:simplfc(3)}) does not admit real solutions
for both of the variables $y_1$ and $y_2$.\/} In other words: {\em In the
case of the simplified model (\ref{eq:simplmod-lagr}), irrespective of the
considered gauge group SU($N$) with $N = 4,8,\dots$, there are no solutions
at all of the one-loop finiteness conditions for Yukawa interaction and
quartic scalar-boson self-couplings, Eqs. (\ref{eq:simplfc(1)}) to
(\ref{eq:simplfc(3)}), with some non-vanishing Yukawa coupling constant $h$,
i.~e., with $h \ne 0$.\/}

The proof of the above statement is based on the (of course, only from the
physical point of view necessary) requirement of reality of our three
variables $x,y_1,y_2$, in particular, of $y_1$. Consider, for any value of
the variable $y_2$, Eq. (\ref{eq:simplfc(2)}) as a quadratic equation for
$y_1$ and obtain the variable $y_1$, as a function of the so far undetermined
variable $y_2$, as one of the two roots of this quadratic equation. Then
reality of $y_1$ is guaranteed if a certain inequality of the generic form
\begin{equation}
a(\ell) \, y_2^2 + b(\ell) \, y_2 + c(\ell) \le 0
\label{eq:ineq-arbitrell}
\end{equation}
is satisfied, where, after some amount of straightforward algebra, for the
non-vanishing solution (\ref{eq:simpl-xsol2}) of Eq. (\ref{eq:simplfc(1)})
the three coefficient functions $a(\ell)$, $b(\ell)$, $c(\ell)$ may be cast
into the form
\begin{eqnarray}
a(\ell)
&=& \dfrac{4 \, \ell^2 - 1}{2 \, \ell^2} \left( 8 \, \ell^2 + 9 \right)
\quad , \nonumber\\[1ex]
b(\ell)
&=& 12 \, \dfrac{4 \, \ell^2 - 1}{184 \, \ell^2 - 3}
\left( 32 \, \ell^2 + 15 \right) \quad , \nonumber\\[1ex]
c(\ell)
&=& 3 \, \dfrac{813056 \, \ell^6 + 346496 \, \ell^4 - 4764 \, \ell^2 - 315}
{\left( 184 \, \ell^2 - 3 \right)^2} \quad . \nonumber
\end{eqnarray}
Recall that the group parameter $\ell$ is necessarily larger than or equal to
1, $\ell \ge 1$. Within this range of $\ell$
\begin{itemize}
\item the coefficient functions $a(\ell)$ and $b(\ell)$ are quite obviously
positive for all values of $\ell$ and,
\item similarly, the coefficient function $c(\ell)$ turns out to be a convex
function of this parameter $\ell$,
$$
\dfrac{d^2}{d\ell^2} \, c(\ell) > 0 \quad \mbox{for} \quad \ell \ge 1 \quad,
$$
and thus to be, in particular, some strictly monotonic increasing function
with increasing $\ell$ and is therefore bounded from below by its value at
$\ell = 1$:
$$
c(\ell) \ge c(1) = \dfrac{3463419}{32761} = 105.71\dots > 0 \quad.
$$
\end{itemize}
Accordingly, all coefficient functions in the inequality
(\ref{eq:ineq-arbitrell}) are strictly positive,
\begin{eqnarray}
a(\ell) &>& 0 \quad \forall\ \ell = 1,2,\dots \quad , \nonumber\\
b(\ell) &>& 0 \quad \forall\ \ell = 1,2,\dots \quad , \nonumber\\
c(\ell) &>& 0 \quad \forall\ \ell = 1,2,\dots \quad . \nonumber
\end{eqnarray}
Therefore, because of the (again only from the physical point of view,
namely, for reasons of stability of the corresponding quantum field theory,
necessary) positivity of the quartic scalar-boson self-couplings, in
particular, of the scaled variable $y_2$, i.~e., $y_2 \ge 0$, the inequality
(\ref{eq:ineq-arbitrell}) cannot be satisfied. This implies that, for the
second, non-vanishing solution for $x$ as given in Eq.
(\ref{eq:simpl-xsol2}), there exists no real solution of the set of equations
(\ref{eq:simplfc(1)}) to (\ref{eq:simplfc(3)}) for the variable $y_1$.

The above findings disprove the claim of the authors of Ref. \cite{shapiro93}
that \ao for $\ell \ge 3$ finite solutions with $h$ as given by Eq.
(\ref{eq:simpl-xsol2}) are also possible." In contrast to this statement, as
has been shown just before, there is {\em no\/} chance to find physically
acceptable, i.~e., real, solutions of the set of equations
(\ref{eq:simplfc(1)}) to (\ref{eq:simplfc(3)}) with $h \ne 0$ for any value
of $\ell$.

In Ref. \cite{lucha93a} there has been demonstrated that within any simplified
model (of the kind introduced in the preceding section) with vanishing Yukawa
coupling constant $h$ the masses of the scalar bosons receive, already at the
one-loop level, quadratically divergent contributions. Therefore, as a
by-product of the result stated above, already at this very early stage we
may ascertain that all theories among the class of simplified models with
vanishing one-loop beta functions, which are to be found as solutions of the
finiteness conditions (\ref{eq:olf-gc/simpl}) and (\ref{eq:simplfc(1)}) to
(\ref{eq:simplfc(3)}), will be plagued by quadratic divergences in the
renormalization of the scalar-boson masses.

As an important consequence of the above introductory statement, it is
sufficient to restrict our discussion to the following reduced set of
equations for $x = 0$:
\begin{eqnarray}
&&\left( 16 \, \ell^2 + 7 \right) y_1^2 + 24 - 48 \, \ell \, y_1
+ \dfrac{16 \, \ell^2 - 4}{\ell}
\left( y_1 \, y_2 + \dfrac{1}{2 \, \ell} \, y_2^2 \right) \nonumber\\
&&\qquad = 0 \quad ,
\label{eq:simplfc(2)x=0}\\[1ex]
&&\dfrac{16 \, \ell^2 - 15}{\ell} \, y_2^2 + 12 \, \ell - 48 \, \ell \, y_2
+ 12 \, y_1 \, y_2 = 0 \quad .
\label{eq:simplfc(3)x=0}
\end{eqnarray}

Our next step is the derivation of both lower as well as upper bounds on the
variables $y_1$ and $y_2$. First of all, both variables $y_1$ and $y_2$ must
be definitely non-vanishing. Examination of the above reduced set of
equations for the two cases $y_1 = 0$ and $y_2 = 0$, respectively, more
precisely, of Eq. (\ref{eq:simplfc(2)x=0}) for $y_1 = 0$,
$$
12 + \dfrac{4 \, \ell^2 - 1}{\ell^2} \, y_2^2 = 0 \quad ,
$$
and of Eq. (\ref{eq:simplfc(3)x=0}) for $y_2 = 0$,
$$
\ell = 0 \quad ,
$$
shows that the reality of the variable $y_2$, on the one hand, and the
requirement $\ell \ne 0$, on the other hand, demand
$$
y_1 \ne 0 \quad \mbox{and} \quad y_2 \ne 0 \quad .
$$
All of the desired bounds on $y_1$ and $y_2$ follow from the existence of a
single negative term on the left-hand side of each of Eqs.
(\ref{eq:simplfc(2)x=0}) and (\ref{eq:simplfc(3)x=0}). Consequently, taking
into account the positivity of the coefficients in front of the various terms
in these two equations, for any valid solution this single negative term has
to counterbalance all positive terms and, in particular, not only the sum of
all of them but also the sum of any subset of them. Investigation of the
terms at most linear in the scaled quartic scalar-boson self-couplings $y_1$
or $y_2$ rather trivially leads, from Eq. (\ref{eq:simplfc(2)x=0}),
$$
1 - 2 \, \ell \, y_1 < 0 \quad ,
$$
to the constraint
$$
y_1 > \dfrac{1}{2 \, \ell}
$$
and, from Eq. (\ref{eq:simplfc(3)x=0}),
$$
(1 - 4 \, y_2 ) \, \ell < 0 \quad ,
$$
to the constraint
$$
y_2 > \dfrac{1}{4} \quad .
$$
Similarly, inclusion of the terms proportional to the product $y_1 \, y_2$ of
the two variables $y_1$, $y_2$ yields, from Eq. (\ref{eq:simplfc(2)x=0}),
$$
6 - 12 \, \ell \, y_1 + \dfrac{4 \, \ell^2 - 1}{\ell} \, y_1 \, y_2 < 0
\quad ,
$$
the constraint
$$
y_2 < \dfrac{6 \, \ell}{4 \, \ell^2 - 1}
\left( 2 \, \ell - \dfrac{1}{y_1} \right)
< \dfrac{12 \, \ell^2}{4 \, \ell^2 - 1} \le 4
$$
and, from Eq. (\ref{eq:simplfc(3)x=0}),
$$
\ell - 4 \, \ell \, y_2 + y_1 \, y_2 < 0 \quad ,
$$
the constraint
$$
y_1 < \ell \left( 4 - \dfrac{1}{y_2} \right) < 4 \, \ell \quad .
$$
Shuffling together all of the above constraints, the numerical values of the
variables $y_1$, $y_2$ which eventually provide some solution to the set of
equations (\ref{eq:simplfc(2)x=0}), (\ref{eq:simplfc(3)x=0}) are unavoidably
restricted to the two ranges
\begin{eqnarray}
\dfrac{1}{2 \, \ell} < &y_1& < 4 \, \ell \quad ,
\label{eq:bounds-y1}\\[1ex]
\dfrac{1}{4} < &y_2& < 4 \quad .
\label{eq:bounds-y2}
\end{eqnarray}
Note that for the variable $y_1$ both bounds depend on $\ell$ whereas for the
variable $y_2$ both bounds do not depend on $\ell$. Rather the latter ones
hold for arbitrary values of the group parameter $\ell$.

By application of our above considerations concerning the reality of the
variable $y_2$ to the case $x = 0$, we are able to show that we may expect to
obtain solutions of the reduced set of equations (\ref{eq:simplfc(2)x=0}),
(\ref{eq:simplfc(3)x=0}) only for values of the group parameter $\ell$ larger
than or equal to 2, i.~e., for $\ell \ge 2$.\footnote{\normalsize\ This
general statement is in accordance with the findings obtained in Ref.
\cite{shapiro93} by some numerical approach for the two special cases $\ell =
1$ and $\ell = 2$.} For the vanishing solution (\ref{eq:simpl-xsol1}) of Eq.
(\ref{eq:simplfc(1)}) the three coefficient functions $a(\ell)$, $b(\ell)$,
$c(\ell)$ in the inequality (\ref{eq:ineq-arbitrell}) read
\begin{eqnarray}
a(\ell)
&=& \dfrac{4 \, \ell^2 - 1}{2 \, \ell^2} \left( 8 \, \ell^2 + 9 \right)
\quad , \nonumber\\[1ex]
b(\ell)
&=& 24 \left( 4 \, \ell^2 - 1 \right) \quad , \nonumber\\[1ex]
c(\ell)
&=& 42 - 48 \, \ell^2 \quad . \nonumber
\end{eqnarray}
Quite obviously, here only the coefficient functions $a(\ell)$ and $b(\ell)$
are positive whereas the coefficient function $c(\ell)$ is negative for
arbitrary $\ell = 1,2,\dots$:
\begin{eqnarray}
a(\ell) &>& 0 \quad \forall\ \ell = 1,2,\dots \quad , \nonumber\\
b(\ell) &>& 0 \quad \forall\ \ell = 1,2,\dots \quad , \nonumber\\
c(\ell) &<& 0 \quad \forall\ \ell = 1,2,\dots \quad . \nonumber
\end{eqnarray}
However, taking into account the lower bound on $y_2$ from
(\ref{eq:bounds-y2}), $y_2 > \frac{1}{4}$, it is easy to convince oneself
that for $x = 0$ the inequality (\ref{eq:ineq-arbitrell}) can only be
satisfied for
$$
\ell^2 > \dfrac{295 + \sqrt{85369}}{368} \simeq 1.6 \quad .
$$
Accordingly, for the value $\ell = 1$ of the group parameter $\ell$ there
exists no solution of the reduced set of equations for $x = 0$, Eqs.
(\ref{eq:simplfc(2)x=0}), (\ref{eq:simplfc(3)x=0}).

Our final step in the course of determination of the complete set of
solutions to the set of equations (\ref{eq:simplfc(1)}) to
(\ref{eq:simplfc(3)}) is the investigation of Eqs. (\ref{eq:simplfc(2)x=0}),
(\ref{eq:simplfc(3)x=0}) in the limit of infinitely large gauge groups,
i.~e., for $\ell \to \infty$. In this limit---which is admittedly hard to
interpret from any physical point of view---the remaining set of finiteness
conditions to be solved, as represented by Eqs. (\ref{eq:simplfc(2)x=0}),
(\ref{eq:simplfc(3)x=0}), simplifies to the, in a certain sense, \ao
asymptotic" set of equations
\begin{eqnarray}
&&2 \, \ell^2 \, y_1^2 + 3 - 6 \, \ell \, y_1 + 2 \, \ell \, y_1 \, y_2
+ y_2^2 = 0 \quad ,
\label{eq:simplfc(2)x=0;linfty}\\[1ex]
&&4 \, \ell \, y_2^2 + 3 \, \ell - 12 \, \ell \, y_2 + 3 \, y_1 \, y_2 = 0
\quad .
\label{eq:simplfc(3)x=0;linfty}
\end{eqnarray}
Now, the first term on the left-hand side of Eq.
(\ref{eq:simplfc(2)x=0;linfty}) is the only one in the above set of
equations which, apart from the a priori unknown dependence of the variable
$y_1$ on $\ell$, is proportional to $\ell^2$. Hence, the only chance for an
eventual balance between the contributions of the various positive and
negative terms in Eq. (\ref{eq:simplfc(2)x=0;linfty}) is that $y_1$ is
of the form of some constant, $k$, divided by the group parameter $\ell$:
$$
y_1 = \dfrac{k}{\ell} \quad .
$$
In this case, Eqs. (\ref{eq:simplfc(2)x=0;linfty}),
(\ref{eq:simplfc(3)x=0;linfty}) read
\begin{eqnarray}
&&2 \, k^2 + 3 - 6 \, k + 2 \, k \, y_2 + y_2^2 = 0 \quad ,
\label{eq:simplfc(2)x=0;y1=k/l}\\[1ex]
&&4 \, y_2^2 + 3 - 12 \, y_2 + 3 \, \dfrac{k}{\ell^2} \, y_2 = 0 \quad .
\label{eq:simplfc(3)x=0;y1=k/l}
\end{eqnarray}
Note that, according to the right-hand inequality in (\ref{eq:bounds-y2}),
the variable $y_2$ is bounded from above by a constant which is independent
from the group parameter $\ell$: $y_2 < \mbox{const}$. For this reason, the
last term on the left-hand side of Eq. (\ref{eq:simplfc(3)x=0;y1=k/l}) does
not contribute in the limit $\ell \to \infty$, that is, it may be dropped in
the limit $\ell \to \infty$. Accordingly, the final form of the asymptotic
set of equations (\ref{eq:simplfc(2)x=0;y1=k/l}),
(\ref{eq:simplfc(3)x=0;y1=k/l}) is
\begin{eqnarray}
&&2 \, k^2 + 3 - 6 \, k + 2 \, k \, y_2 + y_2^2 = 0 \quad ,
\label{eq:simplfc(2)x=0;y1=k/l-final}\\[1ex]
&&4 \, y_2^2 + 3 - 12 \, y_2 = 0 \quad .
\label{eq:simplfc(3)x=0;y1=k/l-final}
\end{eqnarray}
Now the second of our asymptotic set of equations, Eq.
(\ref{eq:simplfc(3)x=0;y1=k/l-final}), does no more depend on the variable
$k$ (or $y_1$, respectively) and therefore may immediately be solved for
$y_2$, with the \ao asymptotic" result
\begin{equation}
y_{2,\infty} = \dfrac{3 \pm \sqrt{6}}{2}
= \left\{ \begin{array}{l} 2.7247\dots \quad , \\ 0.2752\dots \quad .
\end{array} \right.
\label{eq:asympsol-y2}
\end{equation}
Of course, both numerical values of this asymptotic solution lie within the
two bounds on $y_2$ given by Eq. (\ref{eq:bounds-y2}): $\frac{1}{4} <
y_{2,\infty} < 4$. Upon insertion of this result for $y_{2,\infty}$ Eq.
(\ref{eq:simplfc(2)x=0;y1=k/l-final}) becomes, for each of the two solutions
for $y_{2,\infty}$ given in Eq. (\ref{eq:asympsol-y2}), a quadratic equation
for the only up to now unknown variable $k$; the two roots of this quadratic
equation may be written down easily. The necessary reality of the constant
$k$, inherited from the reality of $y_1$, however, is granted exclusively for
the negative sign in front of the square root in Eq. (\ref{eq:asympsol-y2})
and hence only for the lower asymptotic solution for $y_2$, that is, for
$y_{2,\infty} = \frac{1}{2} \left( 3 - \sqrt{6} \right) = 0.2752\dots$. For
this numerical value of $y_{2,\infty}$ the two roots of Eq.
(\ref{eq:simplfc(2)x=0;y1=k/l-final}) are
\begin{equation}
k = \dfrac{3 + \sqrt{6} \pm \sqrt{3 \left( 6 \sqrt{6} - 13 \right)}}{4}
= \left\{ \begin{array}{l} 1.9264\dots \quad , \\ 0.7983\dots \quad .
\end{array} \right.
\label{eq:asympsol-k}
\end{equation}
In summary, the only asymptotic solutions to the set of equations
(\ref{eq:simplfc(1)}) to (\ref{eq:simplfc(3)}) for $\ell \to \infty$ are
given by
\begin{eqnarray}
x &=& 0 \quad ,
\label{eq:asympsol-x-sum}\\[1ex]
y_{1,\infty} &=& \dfrac{3 + \sqrt{6} \pm
\sqrt{3 \left( 6 \sqrt{6} - 13 \right)}}{4 \, \ell}
= \dfrac{1}{\ell} \times
\left\{ \begin{array}{l} 1.9264\dots \quad , \\ 0.7983\dots \quad ,
\end{array} \right.
\label{eq:asympsol-y1-sum}\\[1ex]
y_{2,\infty} &=& \dfrac{3 - \sqrt{6}}{2} = 0.2752\dots \quad .
\label{eq:asympsol-y2-sum}
\end{eqnarray}
One encounters no problems at all in verifying the consistency of the above
asymptotic solution $y_{1,\infty}$ for the variable $y_1$, Eq.
(\ref{eq:asympsol-y1-sum}), with the ($\ell$-dependend) bounds on $y_1$ from
the inequalities (\ref{eq:bounds-y1}): $\frac{1}{2 \, \ell} < y_{1,\infty} <
4 \, \ell$ for both solutions in (\ref{eq:asympsol-y1-sum}) and for arbitrary
$\ell \ge 1$.

{}From Eq. (\ref{eq:asympsol-y2-sum}) we learn that in the limit $\ell \to
\infty$ the solutions to the set of equations (\ref{eq:simplfc(2)x=0}),
(\ref{eq:simplfc(3)x=0}) are degenerate with respect to the variable $y_2$.
For finite $\ell$, however, this degeneracy is removed---for every value of
$\ell$ there exist {\em two but only two\/} different solutions for $y_2$. In
order to see this fact, by expressing the variable $y_1$ from Eq.
(\ref{eq:simplfc(3)x=0}), which is only linear in $y_1$, in terms of $y_2$
and inserting this expression into Eq. (\ref{eq:simplfc(2)x=0}), we consider
the left-hand side of Eq. (\ref{eq:simplfc(2)x=0}) as a (quartic) function of
$y_2$ only and calculate the position of the three extrema of this function.
The relative signs of the values of this function at these extrema then
indicate that it possesses only two real zeros, the two solutions for $y_2$
just mentioned. For each of these solutions the corresponding value of $y_1$
may then be computed unambiguously from Eq. (\ref{eq:simplfc(3)x=0}).

For illustrative purposes, we present in Table \ref{tab:simplsol}, for some
values of the group parameter $\ell$, numerical solutions for the two
variables $y_1$ and $y_2$, as computed by some standard numerical method for
the solution of coupled sets of equations, as well as the corresponding
values of the re-scaled variable $k \equiv y_1 \, \ell$. The observed
behaviour of the solutions for large values of $\ell$ eventually might have
been expected already from the preceding discussion: $y_2$ approaches the
single asymptotic value (\ref{eq:asympsol-y2-sum}) while $k$ approaches the
one or the other of its two possible, constant values (\ref{eq:asympsol-k}).

Summarizing the whole set of findings with respect to the possible solutions
of the set of equations (\ref{eq:simplfc(1)}) to (\ref{eq:simplfc(3)}), the
following simple picture emerges: The complete spectrum of solutions to the
three, Yukawa and quartic scalar-boson self-interaction, finiteness
conditions (\ref{eq:simplfc(1)}) to (\ref{eq:simplfc(3)}) of the simplified
model (\ref{eq:simplmod-lagr})
\begin{itemize}
\item is characterized by a vanishing Yukawa coupling constant $h$, i.~e., by
$$
h = 0 \quad ,
$$
and,
\item for every value of the group parameter $\ell$, consists of precisely
two sequences of solutions for the quartic scalar-boson self-interaction
coupling constants $\lambda_1$, $\lambda_2$ normalized to the square of the
gauge coupling constant $g$, $\lambda_1/g^2$ and $\lambda_2/g^2$, which,
\item starting at $\ell = 2$ with the numerical values
$$
\dfrac{\lambda_1}{g^2} = 0.7762\dots \quad , \qquad
\dfrac{\lambda_2}{g^2} = 0.3027\dots \quad ,
$$
and
$$
\dfrac{\lambda_1}{g^2} = 0.4362\dots \quad , \qquad
\dfrac{\lambda_2}{g^2} = 0.2865\dots \quad ,
$$
respectively,
\item converge for the group parameter $\ell$ increasing beyond any limits
asymptotically towards the behaviour indicated by Eqs.
(\ref{eq:asympsol-y1-sum}) and (\ref{eq:asympsol-y2-sum}): for $\ell \to
\infty$
\begin{eqnarray}
\dfrac{\lambda_1}{g^2} &=& \dfrac{3 + \sqrt{6} \pm
\sqrt{3 \left( 6 \sqrt{6} - 13 \right)}}{4 \, \ell}
= \dfrac{1}{\ell} \times
\left\{ \begin{array}{l} 1.9264\dots \quad , \\ 0.7983\dots \quad ,
\end{array} \right. \nonumber\\[1ex]
\dfrac{\lambda_2}{g^2} &=& \dfrac{3 - \sqrt{6}}{2} = 0.2752\dots
\quad .\nonumber
\end{eqnarray}
\end{itemize}

{\normalsize
\begin{table}[h]
\begin{center}
\caption{Numerical solutions for the two variables $y_1$ and $y_2$ of the
simplified model as well as the re-scaled, asymptotically constant variable
$k \equiv y_1 \, \ell$, all of them for various values of the group parameter
$\ell$. To the precision aimed at here, the asymptotic region \ao$\ell \to
\infty$" is reached already for $\ell \simeq 500$.}\label{tab:simplsol}
\vspace{0.5cm}
\begin{tabular}{|r|ll|l|}
\hline
&&&\\[-1ex]
\multicolumn{1}{|c|}{$\ell$}&\multicolumn{1}{c}{$y_1$}&
\multicolumn{1}{c|}{$y_2$}&\multicolumn{1}{c|}{$k \equiv y_1 \, \ell$}\\
&&&\\[-1.5ex]
\hline\hline
&&&\\[-1.5ex]
$\quad$2$\quad$&$\quad$0.7762\dots$\quad$&$\quad$0.3027\dots$\quad$
&$\quad$1.5524\dots$\quad$\\
&$\quad$0.4362\dots$\quad$&$\quad$0.2865\dots$\quad$
&$\quad$0.8724\dots$\quad$\\
&&&\\[-1.5ex]
\hline
&&&\\[-1.5ex]
$\quad$3$\quad$&$\quad$0.5855\dots$\quad$&$\quad$0.2890\dots$\quad$
&$\quad$1.7565\dots$\quad$\\
&$\quad$0.2754\dots$\quad$&$\quad$0.2798\dots$\quad$
&$\quad$0.8262\dots$\quad$\\
&&&\\[-1.5ex]
\hline
&&&\\[-1.5ex]
$\quad$4$\quad$&$\quad$0.4574\dots$\quad$&$\quad$0.2832\dots$\quad$
&$\quad$1.8297\dots$\quad$\\
&$\quad$0.2033\dots$\quad$&$\quad$0.2777\dots$\quad$
&$\quad$0.8132\dots$\quad$\\
&&&\\[-1.5ex]
\hline
&&&\\[-1.5ex]
$\quad$5$\quad$&$\quad$0.3728\dots$\quad$&$\quad$0.2804\dots$\quad$
&$\quad$1.8641\dots$\quad$\\
&$\quad$0.1615\dots$\quad$&$\quad$0.2768\dots$\quad$
&$\quad$0.8076\dots$\quad$\\
&&&\\[-1.5ex]
\hline
&&&\\[-1.5ex]
$\quad$6$\quad$&$\quad$0.3138\dots$\quad$&$\quad$0.2789\dots$\quad$
&$\quad$1.8830\dots$\quad$\\
&$\quad$0.1341\dots$\quad$&$\quad$0.2763\dots$\quad$
&$\quad$0.8047\dots$\quad$\\
&&&\\[-1.5ex]
\hline
&&&\\[-1.5ex]
$\quad$8$\quad$&$\quad$0.2377\dots$\quad$&$\quad$0.2773\dots$\quad$
&$\quad$1.9019\dots$\quad$\\
&$\quad$0.1002\dots$\quad$&$\quad$0.2758\dots$\quad$
&$\quad$0.8018\dots$\quad$\\
&&&\\[-1.5ex]
\hline
&&&\\[-1.5ex]
$\quad$10$\quad$&$\quad 1.9108\dots\times 10^{-1}\quad$
&$\quad$0.2765\dots$\quad$&$\quad$1.9108\dots$\quad$\\
&$\quad 0.8005\dots\times 10^{-1}\quad$&$\quad$0.2756\dots$\quad$
&$\quad$0.8005\dots$\quad$\\
&&&\\[-1.5ex]
\hline
&&&\\[-1.5ex]
$\quad$50$\quad$&$\quad 3.8516\dots\times 10^{-2}\quad$
&$\quad$0.2753\dots$\quad$&$\quad$1.9258\dots$\quad$\\
&$\quad 1.5968\dots\times 10^{-2}\quad$&$\quad$0.2752\dots$\quad$
&$\quad$0.7984\dots$\quad$\\
&&&\\[-1.5ex]
\hline
&&&\\[-1.5ex]
$\quad$100$\quad$&$\quad 1.9264\dots\times 10^{-2}\quad$
&$\quad$0.2752\dots$\quad$&$\quad$1.9264\dots$\quad$\\
&$\quad 0.7984\dots\times 10^{-2}\quad$&$\quad$0.2752\dots$\quad$
&$\quad$0.7984\dots$\quad$\\
&&&\\[-1.5ex]
\hline
&&&\\[-1.5ex]
$\quad$500$\quad$&$\quad 3.8528\dots\times 10^{-3}\quad$
&$\quad$0.2752\dots$\quad$&$\quad$1.9264\dots$\quad$\\
&$\quad 1.5966\dots\times 10^{-3}\quad$&$\quad$0.2752\dots$\quad$
&$\quad$0.7983\dots$\quad$\\
&&&\\[-1.5ex]
\hline
&&&\\[-1.5ex]
$\quad$1000$\quad$&$\quad 1.9264\dots\times 10^{-3}\quad$
&$\quad$0.2752\dots$\quad$&$\quad$1.9264\dots$\quad$\\
&$\quad 0.7983\dots\times 10^{-3}\quad$&$\quad$0.2752\dots$\quad$
&$\quad$0.7983\dots$\quad$\\
&&&\\[-1.5ex]
\hline
&&&\\[-1.5ex]
$\quad$5000$\quad$&$\quad 3.8528\dots\times 10^{-4}\quad$
&$\quad$0.2752\dots$\quad$&$\quad$1.9264\dots$\quad$\\
&$\quad 1.5966\dots\times 10^{-4}\quad$&$\quad$0.2752\dots$\quad$
&$\quad$0.7983\dots$\quad$\\
&&&\\[-1.5ex]
\hline
&&&\\[-1.5ex]
$\quad$10000$\quad$&$\quad 1.9264\dots\times 10^{-4}\quad$
&$\quad$0.2752\dots$\quad$&$\quad$1.9264\dots$\quad$\\
&$\quad 0.7983\dots\times 10^{-4}\quad$&$\quad$0.2752\dots$\quad$
&$\quad$0.7983\dots$\quad$\\[1.5ex]
\hline
\end{tabular}
\end{center}
\end{table}}

\clearpage

\section{Solutions for the General Model}\label{sec:gensol}

For the general model, the conditions for the one-loop contributions to the
renormalization-group beta functions of the Yukawa couplings $h_i$, $i =
1,2,3$, as well as of the quartic scalar-boson self-couplings $\lambda_i$, $i
= 1,2,\dots ,5$, to vanish \cite{shapiro93}, expressed in terms of the eight
real and non-negative variables
$$
x_i \equiv \dfrac{h_i^2}{g^2} \ge 0 \quad , \qquad i = 1,2,3 \quad ,
$$
and
$$
y_i \equiv \dfrac{\lambda_i}{g^2} \ge 0 \quad , \qquad i = 1,2,\dots ,5
\quad ,
$$
read for arbitrary gauge groups SU($N$):
\begin{eqnarray}
&&4 \, N \, (m + 1) \, x_1^2 - 12 \, N \, x_1 + 2 \, m \, x_1 \, x_2
+ x_1 \, x_3 \nonumber\\
&&\qquad \mp \: \sqrt{x_1 \, x_2} \, x_3 = 0 \quad ,
\label{eq:genfc(1)}\\[1ex]
&&\dfrac{N^2 + 2 \, m \, N - 3}{N} \, x_2^2
- \dfrac{6 \left( N^2 - 1 \right)}{N} \, x_2 + 4 \, m \, N \, x_1 \, x_2
+ \dfrac{N^2 - 1}{N} \, x_2 \, x_3 \nonumber\\
&&\qquad \mp \: 2 \, N \, \sqrt{x_1 \, x_2} \, x_3 = 0 \quad ,
\label{eq:genfc(2)}\\[1ex]
&&\dfrac{(4 \, m + 1) \left( N^2 - 1 \right) + N}{2 \, N} \, x_3^2
- \dfrac{5 \, N^2 + 1}{N} \, x_3 + N \, x_1 \, x_3
+ \dfrac{N^2 - 1}{2 \, N} \, x_2 \, x_3 \nonumber\\
&&\qquad \mp \: 2 \, N \, \sqrt{x_1 \, x_2} \, x_3 = 0 \quad ,
\label{eq:genfc(3)}\\[1ex]
&&\left( N^2 + 7 \right) y_1^2 + 24 - 12 \, N \, y_1
+ \dfrac{4 \left( N^2 - 4 \right)}{N} \, y_2
\left( y_1 + \dfrac{2}{N} \, y_2 \right)
+ 2 \, N \, y_3^2 \nonumber\\
&&\qquad + \: 8 \, m \, N \, x_1 \, y_1 + 4 \, m \, x_2 \, y_1
- 32 \, m \, x_1^2 - \dfrac{4 \, m}{N} \, x_2^2 = 0 \quad ,
\label{eq:genfc(4)}\\[1ex]
&&\dfrac{4 \left( N^2 - 15 \right)}{N} \, y_2^2 - 12 \, N \, y_2
+ 3 \, N + 12 \, y_1 \, y_2 + y_4^2 + 8 \, m \, N \, x_1 \, y_2 \nonumber\\
&&\qquad + \: 4 \, m \, x_2 \, y_2 - 4 \, m \, N \, x_1^2
- 2 \, m \, x_2^2 = 0 \quad ,
\label{eq:genfc(5)}\\[1ex]
&&4 \, y_3^2 - \dfrac{3 \left( 3 \, N^2 - 1 \right)}{N} \, y_3 + 6
+ \left( N^2 + 1 \right) y_1 \, y_3
+ \dfrac{2 \left( N^2 - 4 \right)}{N} \, y_2 \, y_3 \nonumber\\
&&\qquad + \: 2 \, (N + 1) \, y_3 \, y_5
+ \dfrac{2 \left( N^2 - 4 \right)}{N^2} \, y_4^2
+ 4 \, m \, N \, x_1 \, y_3 + 2 \, m \, x_2 \, y_3 \nonumber\\
&&\qquad + \: 2 \, m \, \dfrac{N^2 -1}{N} \, x_3 \, y_3
- 8 \, m \, x_1 \, x_3
- 4 \, m \, \dfrac{N^2 -1}{N^2} \, x_2 \, x_3 \nonumber\\
&&\qquad \pm \: 4 \, m \, \sqrt{x_1 \, x_2} \, x_3 = 0 \quad ,
\label{eq:genfc(6)}\\[1ex]
&&\dfrac{N^2 - 12}{N} \, y_4^2
- \dfrac{3 \left( 3 \, N^2 - 1 \right)}{N} \, y_4 + 3 \, N
+ 2 \, y_1 \, y_4
+ \dfrac{2 \left( N^2 - 8 \right)}{N} \, y_2 \, y_4 \nonumber\\
&&\qquad + \: 2 \, y_4 \, y_5 + 8 \, y_3 \, y_4
+ 4 \, m \, N \, x_1 \, y_4 + 2 \, m \, x_2 \, y_4
+ 2 \, m \, \dfrac{N^2 -1}{N} \, x_3 \, y_4 \nonumber\\
&&\qquad - \: 4 \, m \, N \, x_1 \, x_3
+ \dfrac{4 \, m}{N} \, x_2 \, x_3 = 0 \quad ,
\label{eq:genfc(7)}\\[1ex]
&&2 \, (N + 4) \, y_5^2 - \dfrac{6 \left( N^2 - 1 \right)}{N} \, y_5
+ \left( N^2 -1 \right) y_3^2
+ \dfrac{\left( N^2 - 4 \right) (N - 1)}{2 \, N^2} \, y_4^2 \nonumber\\
&&\qquad + \: \dfrac{3 \, (N - 1) \left( N^2 + 2 \, N - 2 \right)}{2 \, N^2}
+ 4 \, m \, \dfrac{N^2 - 1}{N} \, x_3 \, y_5 \nonumber\\
&&\qquad - \: 2 \, m \,
\dfrac{(N - 1) \left( N^2 + N - 1 \right)}{N^2} \, x_3^2 = 0 \quad .
\label{eq:genfc(8)}
\end{eqnarray}
The indeterminate sign of the terms involving the square root of the
variables $x_1$ and $x_2$ in Eqs. (\ref{eq:genfc(1)}), (\ref{eq:genfc(2)}),
(\ref{eq:genfc(3)}), and (\ref{eq:genfc(6)}) is due to the use of more or
less the squares of the Yukawa coupling constants $h_1$ and $h_2$ as our
basic variables, which forces us to consider both options for this sign in
this kind of expression: $h_i/g = \pm \sqrt{x_i}$ for $i = 1,2,3$.

Unfortunately, the complexity of the set of equations (\ref{eq:genfc(1)}) to
(\ref{eq:genfc(8)}) prevents or at least discourages a thorough analysis
similar to the one that has been performed in the case of the simplified
model. Therefore, instead of attempting a discussion in full generality, we
first focus our attention to the easier to handle special case of vanishing
multiplicity $m$, by setting $m = 0$ in Eqs. (\ref{eq:genfc(1)}) to
(\ref{eq:genfc(8)}).

By closer inspection of the above set of equations (\ref{eq:genfc(1)}) to
(\ref{eq:genfc(8)}), our first observation may be summarized in form of the
statement: {\em In the case of vanishing multiplicity $m$, that is, for $m =
0$, if there exists any solution to the set of finiteness conditions
(\ref{eq:genfc(1)}) to (\ref{eq:genfc(8)}) at all, at least the six different
solutions for the three---\ao Yukawa-interaction-type"---variables
$x_1,x_2,x_3$ to be listed below appear simultaneously with one and the same
set(s) of solutions for the five---\ao self-interaction-type"---variables
$y_1,y_2,y_3,y_4,y_5$, regardless of the value of the integer $N$ which
characterizes the gauge group SU($N$): classifying them with respect to those
among the variables $x_1,x_2,x_3$ which vanish, these solutions read
\begin{enumerate}
\item for all three variables $x_1,x_2,x_3$ vanishing
\begin{equation}
x_1 = x_2 = x_3 = 0 \quad ;
\label{eq:genfc/m=0/sol0}
\end{equation}
\item for both $x_2$ and $x_3$ vanishing
\begin{equation}
x_1 = 3 \quad , \qquad x_2 = x_3 = 0 \quad ;
\label{eq:genfc/m=0/solx1}
\end{equation}
\item for both $x_1$ and $x_3$ vanishing
\begin{equation}
x_2 = \dfrac{6 \left( N^2 - 1 \right)}{N^2 - 3} \quad , \qquad
x_1 = x_3 = 0 \quad ;
\label{eq:genfc/m=0/solx2}
\end{equation}
\item for both $x_1$ and $x_2$ vanishing
\begin{equation}
x_3 = \dfrac{2 \left( 5 \, N^2 + 1 \right)}{N^2 + N - 1} \quad , \qquad
x_1 = x_2 = 0 \quad ;
\label{eq:genfc/m=0/solx3}
\end{equation}
\item for only $x_3$ vanishing
\begin{equation}
x_1 = 3 \quad , \qquad
x_2 = \dfrac{6 \left( N^2 - 1 \right)}{N^2 - 3} \quad , \qquad
x_3 = 0 \quad ;
\label{eq:genfc/m=0/solx1x2}
\end{equation}
and
\item for only $x_2$ vanishing
\begin{eqnarray}
x_1 &=& \dfrac{6 \, N^3 + N^2 - 6 \, N - 1}{N \left( 2 \, N^2 + N - 2 \right)}
\quad , \qquad
x_2 = 0 \quad , \qquad \nonumber\\
x_3 &=& \dfrac{4 \left( 2 \, N^2 + 1 \right)}{2 \, N^2 + N - 2} \quad .
\label{eq:genfc/m=0/solx1x3}
\end{eqnarray}
\end{enumerate}}
\noindent
This observation follows rather trivially from the fact that in the case $m =
0$
\begin{enumerate}
\item there occurs the complete decoupling of Eqs. (\ref{eq:genfc(1)}) to
(\ref{eq:genfc(3)}), which determine the variables $x_1,x_2,x_3$, on the one
hand, from Eqs. (\ref{eq:genfc(4)}) to (\ref{eq:genfc(8)}), which in this
case determine the variables $y_1,y_2,y_3,y_4,y_5$ irrespective of the
numerical values of the variables $x_1,x_2,x_3$, on the other hand, and
\item the set of equations (\ref{eq:genfc(1)}) to (\ref{eq:genfc(3)})---which
in any case determines the variables $x_1,x_2,x_3$---simplifies to
\begin{eqnarray}
&&4 \, N \, x_1^2 - 12 \, N \, x_1 + x_1 \, x_3
\mp \sqrt{x_1 \, x_2} \, x_3 = 0 \quad ,
\label{eq:genfc(1)m=0}\\[1ex]
&&\dfrac{N^2 - 3}{N} \, x_2^2 - \dfrac{6 \left( N^2 - 1 \right)}{N} \, x_2
+ \dfrac{N^2 - 1}{N} \, x_2 \, x_3
\mp 2 \, N \, \sqrt{x_1 \, x_2} \, x_3 \nonumber\\
&&\qquad = 0 \quad ,
\label{eq:genfc(2)m=0}\\[1ex]
&&\dfrac{N^2 + N - 1}{2 \, N} \, x_3^2
- \dfrac{5 \, N^2 + 1}{N} \, x_3 + N \, x_1 \, x_3
+ \dfrac{N^2 - 1}{2 \, N} \, x_2 \, x_3 \nonumber\\
&&\qquad \mp \: 2 \, N \, \sqrt{x_1 \, x_2} \, x_3= 0 \quad .
\label{eq:genfc(3)m=0}
\end{eqnarray}
\end{enumerate}
If any two among the three variables $x_1,x_2,x_3$ vanish, two of the three
equations (\ref{eq:genfc(1)m=0}) to (\ref{eq:genfc(3)m=0}) are identically
satisfied whereas the respective third one reduces to a quadratic equation
for merely that one among these variables which does not vanish a priori. The
two roots of this quadratic equation are then represented either by the
vanishing value of the corresponding variable, which corresponds to the
solution given in (\ref{eq:genfc/m=0/sol0}), or by the value given in Eqs.
(\ref{eq:genfc/m=0/solx1}), (\ref{eq:genfc/m=0/solx2}), or
(\ref{eq:genfc/m=0/solx3}), respectively. If, on the other hand, one and only
one among the three variables $x_1,x_2,x_3$ is required to vanish, one of the
three equations (\ref{eq:genfc(1)m=0}) to (\ref{eq:genfc(3)m=0}) is
identically satisfied. It is then a simple task to extract the values of the
other two, non-vanishing variables from the two remaining equations. For one
of the latter requirements, namely, for $x_1 = 0$, $x_2 \ne 0$, $x_3 \ne 0$,
there arises a conflict with the necessary positivity of our variables
$x_1,x_2,x_3$ whence the corresponding solution does not exist at all.

{}From the above discussion it should have become rather clear that, in order
to decide whether or not there will be solutions to the set of finiteness
conditions (\ref{eq:genfc(1)}) to (\ref{eq:genfc(8)}) for vanishing
multiplicity $m$ at all, we have to investigate, for $m = 0$, the set of
equations which under these circumstances determine the variables
$y_1,y_2,y_3,y_4,y_5$ for their own, namely, Eqs. (\ref{eq:genfc(4)}) to
(\ref{eq:genfc(8)}). In the case $m = 0$ these equations undergo a tremendous
simplification, leaving us with the set of equations
\begin{eqnarray}
&&\left( N^2 + 7 \right) y_1^2 + 24 - 12 \, N \, y_1
+ \dfrac{4 \left( N^2 - 4 \right)}{N} \, y_2
\left( y_1 + \dfrac{2}{N} \, y_2 \right) + 2 \, N \, y_3^2 \nonumber\\
&&\qquad = 0 \quad ,
\label{eq:genfc(4)m=0}\\[1ex]
&&\dfrac{4 \left( N^2 - 15 \right)}{N} \, y_2^2 - 12 \, N \, y_2
+ 3 \, N + 12 \, y_1 \, y_2 + y_4^2 = 0 \quad ,
\label{eq:genfc(5)m=0}\\[1ex]
&&4 \, y_3^2 - \dfrac{3 \left( 3 \, N^2 - 1 \right)}{N} \, y_3 + 6
+ \left( N^2 + 1 \right) y_1 \, y_3
+ \dfrac{2 \left( N^2 - 4 \right)}{N} \, y_2 \, y_3 \nonumber\\
&&\qquad + \: 2 \, (N + 1) \, y_3 \, y_5
+ \dfrac{2 \left( N^2 - 4 \right)}{N^2} \, y_4^2 = 0 \quad ,
\label{eq:genfc(6)m=0}\\[1ex]
&&\dfrac{N^2 - 12}{N} \, y_4^2
- \dfrac{3 \left( 3 \, N^2 - 1 \right)}{N} \, y_4 + 3 \, N
+ 2 \, y_1 \, y_4
+ \dfrac{2 \left( N^2 - 8 \right)}{N} \, y_2 \, y_4 \nonumber\\
&&\qquad + \: 2 \, y_4 \, y_5 + 8 \, y_3 \, y_4 = 0 \quad ,
\label{eq:genfc(7)m=0}\\[1ex]
&&2 \, (N + 4) \, y_5^2 - \dfrac{6 \left( N^2 - 1 \right)}{N} \, y_5
+ \left( N^2 -1 \right) y_3^2
+ \dfrac{\left( N^2 - 4 \right) (N - 1)}{2 \, N^2} \, y_4^2 \nonumber\\
&&\qquad + \: \dfrac{3 \, (N - 1) \left( N^2 + 2 \, N - 2 \right)}{2 \, N^2}
= 0 \quad .
\label{eq:genfc(8)m=0}
\end{eqnarray}
Here, we observe
\begin{itemize}
\item that---in contrast to the most general case as represented by the set
of equations (\ref{eq:genfc(4)}) to (\ref{eq:genfc(8)})---now the left-hand
sides of each of the above relations involve, for $N \ge 5$, just a single
negative term, which has to compensate for all of the positive terms, and
\item that each of these single negative terms is linear in just one of the
variables $y_1,y_2,y_3,y_4,y_5$, each of these variables appearing just once
in one of these negative terms.
\end{itemize}
This observation has two immediate consequences:
\begin{enumerate}
\item All of our five self-interaction-type variables $y_1,y_2,\dots ,y_5$
must be definitely non-vanishing:
$$
y_1 \ne 0 \quad , \quad
y_2 \ne 0 \quad , \quad
y_3 \ne 0 \quad , \quad
y_4 \ne 0 \quad , \quad
y_5 \ne 0 \quad .
$$
If anyone of these variables would vanish, obviously at least one among Eqs.
(\ref{eq:genfc(4)m=0}) to (\ref{eq:genfc(8)m=0}) could not be satisfied,
namely, that one the left-hand side of which contains the negative term
involving this vanishing variable.
\item The obvious necessity of counterbalancing all positive terms on the
left-hand sides of Eqs. (\ref{eq:genfc(4)m=0}) to (\ref{eq:genfc(8)m=0}) by
the single negative term imposes a lot of bounds on the variables
$y_1,y_2,\dots ,y_5$. For $N \ge 5$ the collection of the best of these
bounds reads
\begin{eqnarray}
0 < \dfrac{2}{N}
< &y_1& <
\dfrac{3 \left( 3 \, N^2 - 1 \right)}{N \left( N^2 + 1 \right)}
< \dfrac{9}{N} \le \dfrac{9}{5} \quad ,
\label{eq:bound-y1}\\[1ex]
\dfrac{1}{4}
< &y_2& <
\dfrac{3 \, N^2}{N^2 - 4} \le \dfrac{25}{7} \quad ,
\label{eq:bound-y2}\\[1ex]
0 < \dfrac{2 \, N}{3 \, N^2 - 1}
< &y_3& <
\dfrac{3 \left( 3 \, N^2 - 1 \right)}{8 \, N} < \dfrac{9 \, N}{8} \quad ,
\label{eq:bound-y3}\\[1ex]
\dfrac{1}{3} < \dfrac{N^2}{3 \, N^2 - 1}
< &y_4& <
\dfrac{3 \left( 3 \, N^2 - 1 \right)}{N^2 - 12}
\le \dfrac{222}{13} \quad ,
\label{eq:bound-y4}\\[1ex]
\dfrac{1}{4} < \dfrac{N^2 + 2 \, N - 2}{4 \, N \,(N + 1)}
< &y_5& <
\dfrac{3 \left( N^2 - 1 \right)}{N \, (N + 4)} < 3 \quad .
\label{eq:bound-y5}
\end{eqnarray}
The above bounds may serve to provide some useful guide in any numerical
search for solutions of the set of equations (\ref{eq:genfc(4)m=0}) to
(\ref{eq:genfc(8)m=0}).
\end{enumerate}

The smallest possible gauge group which may serve to illustrate the above
observation is SU($9$): here, for $N =9$ and $m = 0$, the following six
solutions for the three Yukawa-interaction-type variables $x_1,x_2,x_3$,
viz.,
\begin{eqnarray}
&&x_1 = x_2 = x_3 = 0 \quad ,\nonumber\\[1ex]
&&x_1 = 3 \quad , \quad x_2 = x_3 = 0 \quad ,\nonumber\\[1ex]
&&x_2 = \dfrac{80}{13} = 6.1538\dots \x , \quad
x_1 = x_3 = 0 \quad ,\nonumber\\[1ex]
&&x_3 = \dfrac{812}{89} = 9.1235\dots \x , \quad
x_1 = x_2 = 0 \quad ,\nonumber\\[1ex]
&&x_1 = 3 \quad , \quad x_2 = \dfrac{80}{13} = 6.1538\dots \x , \quad
x_3 = 0 \quad ,\nonumber\\[1ex]
&&x_1 = \dfrac{4400}{1521} = 2.8928\dots \x , \quad
x_2 = 0 \quad , \quad x_3 = \dfrac{652}{169} = 3.8579\dots \x ,\nonumber
\end{eqnarray}
accompany both of the two sets of only numerically found solutions for the
five self-interaction-type variables $y_1,y_2,y_3,y_4,y_5$ listed in Table
\ref{tab:gensol-N=9,m=0}. Of course, both of these sets of numerical
solutions respect the bounds on the variables $y_1,y_2,\dots ,y_5$ given in
Eqs. (\ref{eq:bound-y1}) to (\ref{eq:bound-y5}).

{\normalsize
\begin{table}[h]
\begin{center}
\caption{Numerical solutions for the five self-interaction-type variables
$y_1,y_2,y_3,y_4,y_5$ of the general model with gauge group SU($9$) and
multiplicity $m = 0$.}\label{tab:gensol-N=9,m=0}
\vspace{0.5cm}
\begin{tabular}{|c|l|l|}
\hline
&&\\[-1ex]
\multicolumn{1}{|c|}{Variable}&\multicolumn{1}{c|}{Solution I}&
\multicolumn{1}{c|}{Solution II}\\
&&\\[-1.5ex]
\hline
&&\\[-1.5ex]
$\quad y_1\quad$&$\quad$0.4017\dots$\quad$&$\quad$0.5054\dots$\quad$\\
$\quad y_2\quad$&$\quad$0.2864\dots$\quad$&$\quad$0.2907\dots$\quad$\\
$\quad y_3\quad$&$\quad$0.1862\dots$\quad$&$\quad$0.3676\dots$\quad$\\
$\quad y_4\quad$&$\quad$0.3860\dots$\quad$&$\quad$0.4008\dots$\quad$\\
$\quad y_5\quad$&$\quad$0.4167\dots$\quad$&$\quad$0.7812\dots$\quad$\\[1.5ex]
\hline
\end{tabular}
\end{center}
\end{table}}

The mere existence of solutions to the finiteness conditions
(\ref{eq:genfc(1)}) to (\ref{eq:genfc(8)}) for $N = 9$ and $m = 0$ clearly
contradicts and therefore corrects the impression one might get from Ref.
\cite{shapiro93} that solutions to this set of equations with $h_1 = h_2 =
h_3 = 0$ do only exist for $N \ge 13$ but not for $N = 9$. While we find
indeed no solutions to the set of equations (\ref{eq:genfc(4)m=0}) to
(\ref{eq:genfc(8)m=0})---which describes the situation realized for $h_1 =
h_2 = h_3 = 0$---for $N = 5$, Table \ref{tab:gensol-N=9,m=0} presents the
corresponding sets of solutions for $N = 9$.

We would like to address the question of the eventual presence of quadratic
divergences in the scalar-boson masses also for the class of general models.
The relevant analytic expressions for the quadratically divergent one-loop
contributions to the masses of both types of scalar bosons in the general
model, $\Phi$ and $\varphi$, have been calculated already in Ref.
\cite{lucha93a}. Apart from some trivial factors, these quadratic divergences
turn out to be proportional to some quantities $Q$ involving the various
coupling constants in the theory. Expressed in terms of our variables $x_i$,
$i = 1,2,3$, and $y_i$, $i = 1,2,\dots ,5$, the quantities $Q$ read in the
sector of the scalar bosons $\Phi$ \cite{lucha93a}
\begin{eqnarray}
\dfrac{1}{g^2} \, Q^{(\Phi)}
&=& 6 \, N - 4 \, m \, (2 \, N \, x_1 + x_2) \nonumber\\
&+& \left(N^2 + 1\right) y_1 + 2 \, \dfrac{N^2 - 4}{N} \, y_2 + 2 \, N \, y_3
\label{eq:QS-Phi/g2}
\end{eqnarray}
and in the sector of the scalar bosons $\varphi$ \cite{lucha93a}
\begin{eqnarray}
\dfrac{1}{g^2} \, Q^{(\varphi)}
&=& 3 \, \dfrac{N^2 - 1}{N} - 4 \, m \, \dfrac{N^2 - 1}{N} \, x_3 \nonumber\\
&+& \left(N^2 - 1\right) y_3 + 2 \, (N + 1) \, y_5 \quad .
\label{eq:QS-varphi/g2}
\end{eqnarray}
For vanishing multiplicity $m$ all of the various contributions on the
right-hand sides of Eqs. (\ref{eq:QS-Phi/g2}) and (\ref{eq:QS-varphi/g2}) are
strictly positive. This implies that in this case both of the quantities
$Q^{(\Phi)}$ and $Q^{(\varphi)}$ are necessarily non-vanishing. Hence, we are
forced to conclude that for $m = 0$ one will encounter quadratic divergences
in the course of renormalization of scalar-boson masses in every single one
among the class of general models, irrespective of the precise numerical
values of the involved coupling constants $h_i$, $i = 1,2,3$, and
$\lambda_i$, $i = 1,2,3,5$.

{\normalsize
\begin{table}[hbt]
\begin{center}
\caption{Numerical solutions for the three Yukawa-interaction-type variables
$x_1,x_2,x_3$ and the five self-interaction-type variables
$y_1,y_2,y_3,y_4,y_5$ of the general model with gauge group SU($5$) and
multiplicity $m = 1$.}\label{tab:gensol-N=5,m=1}
\vspace{0.5cm}
\begin{tabular}{|c|l|l|}
\hline
&&\\[-1ex]
\multicolumn{1}{|c|}{Variable}&\multicolumn{1}{c|}{Solution I}&
\multicolumn{1}{c|}{Solution II}\\
&&\\[-1.5ex]
\hline
&&\\[-1.5ex]
$\quad x_1\quad$&$\quad\x$1.4642\dots$\quad$&$\quad\x$1.4211\dots$\quad$\\
$\quad x_2\quad$&$\quad\x$0$\quad$&$\quad\x$1.6806\dots$\quad$\\
$\quad x_3\quad$&$\quad\x$1.4303\dots$\quad$&$\quad\x$2.3612\dots$\quad$\\
[1ex]
$\quad y_1\quad$&$\quad\x$0.7059\dots$\quad$&$\quad\x$0.6594\dots$\quad$\\
$\quad y_2\quad$&$\quad\x$1.3872\dots$\quad$&$\quad\x$1.2933\dots$\quad$\\
$\quad y_3\quad$&$\quad\x$0.1692\dots$\quad$&$\quad\x$0.3235\dots$\quad$\\
$\quad y_4\quad$&$\quad\x$1.6480\dots$\quad$&$\quad\x$1.6765\dots$\quad$\\
$\quad y_5\quad$&$\quad\x$0.6067\dots$\quad$&$\quad\x$1.0385\dots$\quad$\\
&&\\[-1.5ex]
\hline
&&\\[-1.5ex]
$\quad Q^{(\Phi)}\quad$&$\quad\x$3.129\dots$\quad$&$\x\: -2.321\dots\quad$\\
$\quad Q^{(\varphi)}\quad$&$\x\: -1.719\dots\quad$&$\, -10.708\dots\quad$\\
[1.5ex]
\hline
\end{tabular}
\end{center}
\end{table}}

{\normalsize
\begin{table}[hbt]
\begin{center}
\caption{Numerical solutions for the three Yukawa-interaction-type variables
$x_1,x_2,x_3$ and the five self-interaction-type variables
$y_1,y_2,y_3,y_4,y_5$ of the general model with gauge group SU($5$) and
multiplicity $m = 2$.}\label{tab:gensol-N=5,m=2}
\vspace{0.5cm}
\begin{tabular}{|c|l|l|}
\hline
&&\\[-1ex]
\multicolumn{1}{|c|}{Variable}&\multicolumn{1}{c|}{Solution I}&
\multicolumn{1}{c|}{Solution II}\\
&&\\[-1.5ex]
\hline
&&\\[-1.5ex]
$\quad x_1\quad$&$\quad\x$0.9847\dots$\quad$&$\quad\x$0.9678\dots$\quad$\\
$\quad x_2\quad$&$\quad\x$0$\quad$&$\quad\x$0.3725\dots$\quad$\\
$\quad x_3\quad$&$\quad\x$0.9174\dots$\quad$&$\quad\x$1.1525\dots$\quad$\\
[1ex]
$\quad y_1\quad$&$\quad\x$0.6627\dots$\quad$&$\quad\x$0.6357\dots$\quad$\\
$\quad y_2\quad$&$\quad\x$0.7087\dots$\quad$&$\quad\x$0.6606\dots$\quad$\\
$\quad y_3\quad$&$\quad\x$0.1709\dots$\quad$&$\quad\x$0.1668\dots$\quad$\\
$\quad y_4\quad$&$\quad\x$0.9075\dots$\quad$&$\quad\x$1.0056\dots$\quad$\\
$\quad y_5\quad$&$\quad\x$0.4079\dots$\quad$&$\quad\x$0.5618\dots$\quad$\\
&&\\[-1.5ex]
\hline
&&\\[-1.5ex]
$\quad Q^{(\Phi)}\quad$&$\, -23.882\dots\quad$&$\, -26.663\dots\quad$\\
$\quad Q^{(\varphi)}\quad$&$\, -11.833\dots\quad$&$\, -19.112\dots\quad$\\
[1.5ex]
\hline
\end{tabular}
\end{center}
\end{table}}

{\normalsize
\begin{table}[hbt]
\begin{center}
\caption{Numerical solutions for the three Yukawa-interaction-type variables
$x_1,x_2,x_3$ and the five self-interaction-type variables
$y_1,y_2,y_3,y_4,y_5$ of the general model with gauge group SU($9$) and
multiplicity $m = 1$.}\label{tab:gensol-N=9,m=1}
\vspace{0.5cm}
\begin{tabular}{|c|l|l|l|l|}
\hline
&&&&\\[-1ex]
\multicolumn{1}{|c|}{Variable}&\multicolumn{1}{c|}{Solution I}&
\multicolumn{1}{c|}{Solution II}&\multicolumn{1}{c|}{Solution III}&
\multicolumn{1}{c|}{Solution IV}\\
&&&&\\[-1.5ex]
\hline
&&&&\\[-1.5ex]
$\quad x_1\quad$&$\quad\x$0$\quad$&$\quad\x$0$\quad$
&$\quad\x$1.4805\dots$\quad$&$\quad\x$1.4546\dots$\quad$\\
$\quad x_2\quad$&$\quad\x$0$\quad$&$\quad\x$0$\quad$&$\quad\x$0$\quad$
&$\quad\x$1.7417\dots$\quad$\\
$\quad x_3\quad$&$\quad\x$0$\quad$&$\quad\x$0$\quad$
&$\quad\x$1.3988\dots$\quad$&$\quad\x$2.3294\dots$\quad$\\[1ex]
$\quad y_1\quad$&$\quad\x$0.4017\dots$\quad$&$\quad\x$0.5054\dots$\quad$
&$\quad\x$0.4336\dots$\quad$&$\quad\x$0.4149\dots$\quad$\\
$\quad y_2\quad$&$\quad\x$0.2864\dots$\quad$&$\quad\x$0.2907\dots$\quad$
&$\quad\x$1.2385\dots$\quad$&$\quad\x$1.1947\dots$\quad$\\
$\quad y_3\quad$&$\quad\x$0.1862\dots$\quad$&$\quad\x$0.3676\dots$\quad$
&$\quad\x$0.0923\dots$\quad$&$\quad\x$0.1756\dots$\quad$\\
$\quad y_4\quad$&$\quad\x$0.3860\dots$\quad$&$\quad\x$0.4008\dots$\quad$
&$\quad\x$1.4849\dots$\quad$&$\quad\x$1.7329\dots$\quad$\\
$\quad y_5\quad$&$\quad\x$0.4167\dots$\quad$&$\quad\x$0.7812\dots$\quad$
&$\quad\x$0.7222\dots$\quad$&$\quad\x$1.1369\dots$\quad$\\
&&&&\\[-1.5ex]
\hline
&&&&\\[-1.5ex]
$\quad Q^{(\Phi)}\quad$&$\x\;\, 95.201\dots\quad$&$\x 107.041\dots\quad$
&$\quad\x 5.816\dots\quad$&$\x\: -0.074\dots\quad$\\
$\quad Q^{(\varphi)}\quad$&$\x\;\, 49.901\dots\quad$&$\x\;\, 71.706\dots\quad$
&$\x\: -1.241\dots\quad$&$\, -19.369\dots\quad$\\[1.5ex]
\hline
\end{tabular}
\end{center}
\end{table}}

{\normalsize
\begin{table}[hbt]
\begin{center}
\caption{Numerical solutions for the three Yukawa-interaction-type variables
$x_1,x_2,x_3$ and the five self-interaction-type variables
$y_1,y_2,y_3,y_4,y_5$ of the general model with gauge group SU($9$) and
multiplicity $m = 2$.}\label{tab:gensol-N=9,m=2}
\vspace{0.5cm}
\begin{tabular}{|c|l|l|l|l|}
\hline
&&&&\\[-1ex]
\multicolumn{1}{|c|}{Variable}&\multicolumn{1}{c|}{Solution I}&
\multicolumn{1}{c|}{Solution II}&\multicolumn{1}{c|}{Solution III}&
\multicolumn{1}{c|}{Solution IV}\\
&&&&\\[-1.5ex]
\hline
&&&&\\[-1.5ex]
$\quad x_1\quad$&$\quad\x$0$\quad$&$\quad\x$0$\quad$
&$\quad\x$0.9917\dots$\quad$&$\quad\x$0.9817\dots$\quad$\\
$\quad x_2\quad$&$\quad\x$0$\quad$&$\quad\x$0$\quad$&$\quad\x$0$\quad$
&$\quad\x$0.3878\dots$\quad$\\
$\quad x_3\quad$&$\quad\x$0$\quad$&$\quad\x$0$\quad$
&$\quad\x$0.8934\dots$\quad$&$\quad\x$1.1273\dots$\quad$\\[1ex]
$\quad y_1\quad$&$\quad\x$0.4017\dots$\quad$&$\quad\x$0.5054\dots$\quad$
&$\quad\x$0.3775\dots$\quad$&$\quad\x$0.3685\dots$\quad$\\
$\quad y_2\quad$&$\quad\x$0.2864\dots$\quad$&$\quad\x$0.2907\dots$\quad$
&$\quad\x$0.7145\dots$\quad$&$\quad\x$0.6880\dots$\quad$\\
$\quad y_3\quad$&$\quad\x$0.1862\dots$\quad$&$\quad\x$0.3676\dots$\quad$
&$\quad\x$0.0901\dots$\quad$&$\quad\x$0.0899\dots$\quad$\\
$\quad y_4\quad$&$\quad\x$0.3860\dots$\quad$&$\quad\x$0.4008\dots$\quad$
&$\quad\x$0.8547\dots$\quad$&$\quad\x$0.9858\dots$\quad$\\
$\quad y_5\quad$&$\quad\x$0.4167\dots$\quad$&$\quad\x$0.7812\dots$\quad$
&$\quad\x$0.4621\dots$\quad$&$\quad\x$0.6088\dots$\quad$\\
&&&&\\[-1.5ex]
\hline
&&&&\\[-1.5ex]
$\quad Q^{(\Phi)}\quad$&$\x\;\, 95.201\dots\quad$&$\x 107.041\dots\quad$
&$\, -44.001\dots\quad$&$\, -46.862\dots\quad$\\
$\quad Q^{(\varphi)}\quad$&$\x\;\, 49.901\dots\quad$&$\x\;\, 71.706\dots\quad$
&$\, -20.412\dots\quad$&$\, -34.131\dots\quad$\\[1.5ex]
\hline
\end{tabular}
\end{center}
\end{table}}

For non-vanishing multiplicity $m$ it is hard to make any general statements
concerning the spectrum of solutions to the set of finiteness conditions
(\ref{eq:genfc(1)}) to (\ref{eq:genfc(8)}) based on purely algebraic
investigations. It is, however, a rather simple task to determine the
corresponding solutions by some numerical method. Here, we merely intend to
exemplify this particular state of affairs by reporting in Tables
\ref{tab:gensol-N=5,m=1} to \ref{tab:gensol-N=9,m=2}, for both of the two
non-vanishing multiplicities $m$ allowed according to Subsection
\ref{subsec:genmod} by one-loop finiteness of the gauge coupling constant
renormalization, $m = 1$ and $m = 2$, all the sets of solutions obtained
numerically within the two smallest possible gauge groups, that is, SU($5$)
and SU($9$), for the three Yukawa-interaction-type variables $x_1,x_2,x_3$ as
well as for the five self-interaction-type variables $y_1,y_2,y_3,y_4,y_5$.
Some but not all of these solutions have already been found in Ref.
\cite{shapiro93}. In fact, in Ref. \cite{shapiro93} exactly one set of
solutions per gauge group and multiplicity has been given.

In the case of vanishing Yukawa-interaction-type variables $x_1,x_2,x_3$ the
set of equations (\ref{eq:genfc(1)}) to (\ref{eq:genfc(8)}) becomes
completely independent from the multiplicity $m$: in this case Eqs.
(\ref{eq:genfc(1)}) to (\ref{eq:genfc(3)}) are satisfied identically whereas
Eqs. (\ref{eq:genfc(4)}) to (\ref{eq:genfc(8)}) reduce to the simplified set
of equations (43) to (47). Consequently, for some given gauge group SU($N$),
that is, for a definite choice of the value of $N$, one and the same sets of
solutions with $x_1 = x_2 = x_3 = 0$ must appear for all conceivable
multiplicities $m = 0,1,2$. And indeed, by comparing the two solutions given
in Table \ref{tab:gensol-N=9,m=0} with Solution I and II in Table
\ref{tab:gensol-N=9,m=1} as well as in Table \ref{tab:gensol-N=9,m=2},
respectively, the attentive reader will find out, at least for the instance
of the gauge group SU($9$), that this duplication of those solutions where
all Yukawa coupling constants vanish, i.~e., of the solutions with $h_1 = h_2
= h_3 = 0$, really takes place.

Let us bring up the question of quadratic divergences for the very last time.
Computing the quantities $Q^{(\Phi)}$ and $Q^{(\varphi)}$ from Eqs.
(\ref{eq:QS-Phi/g2}) and (\ref{eq:QS-varphi/g2}) for all sets of numerical
solutions for the Yukawa-interaction and scalar-boson self-interaction
coupling constants listed in Tables \ref{tab:gensol-N=5,m=1} to
\ref{tab:gensol-N=9,m=2}, we find the corresponding values quoted also in
these tables. Since all of these values are definitely non-vanishing, we end
up with the insight that in each of these examples for general one-loop
finite models with non-vanishing multiplicity $m$ there will arise quadratic
divergences for the masses of the scalar bosons.

\section{Summary and Conclusions}\label{sec:solconc}

The present discussion has been devoted to the explicit construction of two
particular classes---or, more precisely, sequences---of grand unified
theories singled out from the most general case by the requirement of
vanishing one-loop contributions to all the renormalization-group beta
functions of the dimensionless coupling constants in the theory. The particle
content of both of these two sequences of models is restricted to transform
according to some reducible representation of the gauge group which is
(equivalent to) the direct sum of (certain multiples of) the fundamental and
the adjoint representations of the gauge group.

For both types of models, that is, the more or less rather simple one as well
as the slightly more sophisticated one, we intended to extract in a totally
algebraic way from the one-loop finiteness conditions for the Yukawa and
scalar-boson self-interaction coupling constants as much information as
possible about the spectrum of theories fulfilling these requirements. In the
case of the class of simplified models, because of, on the one hand, the
rather high degree of simplicity of this class of models and, on the other
hand, the existence, as a consequence of the finiteness of the
renormalization of the gauge coupling constant at the one-loop level, of a
unique relationship between the gauge group and the otherwise arbitrary
multiplicity of the fermionic particle content, we were able to draw a
coherent picture. Herein we clearly discern some sort of \ao ladder with
converging uprights," which is formed by precisely two distinct solutions for
each valid gauge group. In the evidently more complicated case of the class
of general models we found it advisable to distinguish between the
conceivable values of the relevant multiplicity of fermion representations
contained in the above-mentioned reducible representation and, in particular,
to investigate theories with vanishing or either of the two admissible
non-vanishing multiplicities separately. For vanishing multiplicity we worked
out, in a very similar manner as in the case of the class of simplified
models, the most essential features of this subset of models. For
non-vanishing multiplicity we simply gave the explicit sets of grand unified
theories which correspond to the two smallest possible and therefore perhaps
most interesting gauge groups.

Since all the models in the above two classes of theories have been subjected
just to the requirement of vanishing one-loop beta functions, we have been
interested, of course, in the question whether or not there arise quadratic
divergences. Taking advantage from the fact that, for both classes of
theories, the explicit expressions for the quadratically divergent one-loop
contributions to all the scalar-boson masses may be read off immediately from
Ref. \cite{lucha93a}, we are led to the conclusion that the scalar bosons of
all theories within the complete class of simplified models as well as both
types of scalar bosons within the few explicitly constructed examples of
models of the more general kind receive, not very surprisingly, already at
the one-loop level quadratically divergent contributions to their masses.
Furthermore, as far as the masses of the gauge vector bosons are concerned,
it has been demonstrated already in Ref. \cite{lucha93a} that, within both
classes of theories, for the vector-boson masses the same statement as above
on the occurrence of quadratic divergences holds in any case.

Needless to say that all our statements based on one or the other algebraic
argument may be and have been checked by a corresponding solution of the set
of equations under consideration by some numerical methods.

\end{document}